\documentclass[proceedings]{JHEP3}
\usepackage{amssymb}
\usepackage{amsfonts}
\usepackage{amsmath}
\usepackage{epsfig,multicol}

\newbox\mybox

\newcommand\fverb{\setbox\mybox=\hbox\bgroup\verb}
\newcommand\fverbdo{\egroup\medskip\noindent\fbox{\unhbox\mybox}\ }
\newcommand\fverbit{\egroup\item[\fbox{\unhbox\mybox}]}
\conference{Exactly solvable potentials for q-deformed Coxeter groups}
\abstract{We establish that by parameterizing the configuration space of a one-dimensional
quantum system by polynomial invariants of q-deformed Coxeter groups it is possible
to construct exactly solvable models of Calogero type. We adopt the previously introduced notion
of solvability which consists of relating the Hamiltonian to finite dimensional representation 
spaces of a Lie algebra. We present explicitly the $G_2^q $-case for which we construct 
the potentials by means of suitable gauge transformations.}

\title{Exactly solvable potentials of Calogero type for q-deformed Coxeter
groups}
\author{Andreas Fring$^\bullet$ and Christian Korff$^\circ$ \\
$^\bullet$  Centre for Mathematical Science, City University, \\
$\;$ Northampton Square, London EC1V 0HB, UK\\
$\;$ E-mail: \email{A.Fring@city.ac.uk}\\
$^\circ$ School of Mathematics, University of Edinburgh, \\
$\;$ Mayfield Road, Edinburgh EH9 3JZ, UK\\
$\;$ E-mail: \email{C.Korff@ed.ac.uk}}

\input{tcilatex}

\begin{document}

\section{Introduction}

One of the ultimate goals in the study of quantum mechanical systems is to
find explicit and possibly exact solutions for the eigensystem of
Hamiltonian systems. The Calogero \cite{Cal1,Cal2,Cal3} and Sutherland \cite%
{Suth1,Suth2,Suth3,Suth4} models are some of the well known examples for
theories which are integrable and can be solved exactly, classically as well
as quantum mechanically. The integrability of the models was established
more systematically by relating them to Lie algebraic structures, in the
so-called Hamiltonian reduction method \cite{OP1,OP2,OP3} or by formulating
Lax pairs and zero curvature conditions \cite{DIO,Sas6,Sas5,Sas4,Sas3,Sas2}.
Relatively recent \cite{Tur0,RT,Cap,TurF4} the procedure to establish their
exact solvability (which is conceptionally different from integrability) was
put on a more systematic ground by relating first the coordinates of the
configuration space of the Hamiltonians to invariant polynomials. It was
shown that the differential operators in these polynomials form a
representation for certain algebras, albeit not uniquely. Having an
algebraic version of the model, solvability can be established thereafter by
noting that the eigenfunctions form a flag which coincides with the finite
dimensional representation space of a gl(N)-Lie algebra. This approach has
turned out to be successful in many cases and could even be extended to
theories which are supersymmetric \cite{Brink}.

In a sequence of publications \cite{HR1,HR2,HR3,HR4} this procedure was
reversed. Instead of starting with a concrete potential for a theory,
Haschke and R\"{u}hl proposed to start with a Hamiltonian already formulated
in terms of invariant polynomials and construct the potential from it.
Hence, in this approach the solvability is already build in from the very
start and the question is addressed if possibly one obtains new types of
potentials which are related to solvable models, which are potentially also
integrable. For several examples of models formulated in terms of invariants
of the Weyl Group \cite{HR1,HR2,HR3} and even for Coxeter groups which are
not Weyl groups \cite{HR4,Reidun} it was shown that this is indeed possible.

The main purpose of this paper is to demonstrate that this procedure can
also be carried out successfully for models which are related to q-deformed
Coxeter groups. We demonstrate this for $G_{2}^{q}$. At the same time we
show that also for these groups the associated Hamiltonians can be
formulated in terms of the gl(N)-Lie algebra generators, hence guaranteeing
their solvability.

We shall focus here mainly on the construction of new potentials of Calogero
type, adopting to a large extent the point of view of the aforementioned
papers. The obvious question of solving the associated Schr\"{o}dinger
problem similar as it has been done for the $A_N$ case in \cite%
{Cal1,Cal2,Cal3} shall not be our concern here. While this is an interesting
problem for future work, it appears that it is still open even for almost
all of the non-deformed Lie algebras other than the $A_{N}$-series.

Our manuscript is organized as follows: In the next section we recall the
notion of solvability based on the fact that certain types of Hamiltonians
can be formulated in terms of the generators of the Borel subalgebra of the
gl(N)-Lie algebra. We show how from this formulation one may systematically
construct potentials. In section 3 we assemble the main mathematical
properties about polynomial invariants of the Coxeter group, which play the
crucial role of coordinates in this context. In section 4 we extend these
ideas to the q-deformed Coxeter groups. In section 5 we discuss how certain
choices of the pre-potential lead to Cologero type potentials. In sections 6
and 7 we discuss the Calogero model for $G_{2}$ and its q-deformed version,
respectively, deriving some explicit Calogero type potentials. We state our
conclusions in section 8.

\section{Construction of exactly solvable potentials}

We start by recalling the notion of exact solvability as proposed originally
by A. Turbiner \cite{Tur0} about ten years ago. For this we require
polynomial spaces of the form 
\begin{equation}
V_{n}=\limfunc{span}\left\{ I_{2}^{k_{2}}I_{3}^{k_{3}}\ldots
I_{N}^{k_{N}}\left\vert \sum\nolimits_{i=2}^{N}k_{i}=n\right. \right\} ~.
\end{equation}%
The $I_{i}$ constitute some generic set of variables not further specified
at this point. Evidently, these spaces are embedded into each other $%
V_{0}\subset V_{1}\subset V_{2}\subset \ldots $, hence forming an infinite
flag. A Hamiltonian operator $\mathcal{H}$ acting on such spaces and
respecting 
\begin{equation}
\mathcal{H}:~V_{n}\mapsto V_{n}  \label{H}
\end{equation}%
possesses an infinite family of polynomial eigenfunctions. Therefore, it is
natural to refer to such type of Hamiltonians as \textit{exactly solvable}.

It is now a matter of identifying the spaces $V_{n}$, which of course allows
for numerous solutions. It was noticed that many known models can be fitted
into this scheme when one identifies $V_{n}$ with a finite dimensional
representation space of a gl(N)-Lie algebra. A simple representation of this
algebra in terms of first order differential operators is found when
expressing the usual gl(N)-generators $E_{ij}$ as 
\begin{equation}
E_{ij}\equiv J_{ij}^{0}=I_{i}\partial _{j},\quad E_{i0}\equiv
J_{i}^{-}=\partial _{j},\quad E_{0i}\equiv J_{i}^{+}=\kappa
I_{i}-\sum\limits_{k=2}^{N}I_{i}I_{k}\partial _{k}\quad \text{for }\kappa
\in \mathbb{R}^{+},  \label{J}
\end{equation}%
where $\kappa $ is an arbitrary constant and $\partial _{i}=\partial
/\partial I_{i}$. It is easy to check that these differential operators
satisfy indeed the usual gl(N)-commutation relations 
\begin{equation}
\left[ E_{ij},E_{kl}\right] =\delta _{jk}E_{il}-\delta _{li}E_{kj}~.
\end{equation}%
According to the representation (\ref{J}) all Hamiltonians which are
expressible in terms of the Borel subalgebra of gl(N), i.e. involving only
the generators $J^{0}$ and $J^{-}$, will respect (\ref{H}) and are therefore
exactly solvable. Remarkably, it has turned out that many known solvable
models of Calogero and Sutherland type can be brought into the general form 
\begin{equation}
\mathcal{H=}\sum c_{ij}^{kl}J_{ij}^{0}J_{kl}^{0}+\sum \tilde{c}%
_{ij}^{k}J_{ij}^{0}J_{k}^{-}+\sum \hat{c}_{i}J_{i}^{-}+\sum \check{c}%
_{ij}J_{ij}^{0}~,  \label{HJ}
\end{equation}%
with $c_{ij}^{kl}$, $\tilde{c}_{ij}^{k}$, $\hat{c}_{i}$, $\check{c}_{ij}\in 
\mathbb{R}$ being some coupling constants.

Unfortunately not all models, in particular the ones we shall discuss below,
can be fitted into the gl(N)-framework. Nonetheless, following the same
ideology as outlined above, one can appeal to some other algebras which can
be realized with different types of differential operators than those
provided in (\ref{J}). For our purposes the semi-direct sum gl$_{2}(\mathbb{R%
})\ltimes \mathbb{R}^{\ell +1}$ will be rather useful \cite{Tur0}. It may be
realised by the $\ell +5$ generators%
\begin{eqnarray}
J^{1} &=&\partial _{1},\quad \quad J^{2}=I_{1}\partial _{1}-\frac{\kappa }{3}%
,\quad \quad J^{3}=I_{2}\partial _{2}-\frac{\kappa }{3\ell },\quad
\label{j1} \\
J^{4} &=&I_{1}^{2}\partial _{1}+\ell I_{1}I_{2}\partial _{2}-\kappa
I_{1},\quad \quad J^{5+i}=I_{1}^{i}\partial _{2}~~~~\qquad \text{for }1\leq
i\leq \ell ,~\kappa \in \mathbb{R}^{+}~.  \label{j2}
\end{eqnarray}%
The further condition $\kappa \in \mathbb{Z}^{+}$ guarantees that the
representation is finite dimensional. Expressing now Hamiltonians in terms
of the Borel subalgebra of gl$_{2}(\mathbb{R})\ltimes \mathbb{R}^{\ell +1}$,
i.e. the $J^{i}$ for $1\leq i\leq \ell +5$ with $i\neq 4$, the flag space of
the form%
\begin{equation}
\bar{V}_{n}=\limfunc{span}\left\{ I_{1}^{k_{1}}I_{2}^{k_{2}}\left\vert 0\leq
k_{1}+\ell k_{2}\leq n\right. \right\}
\end{equation}%
will be left invariant in the sense (\ref{H}).

The above observations inspired the starting point of the approach in \cite%
{HR1,HR2,HR3,HR4} which is the eigenvalue equation for the function $\varphi
(\vec{I})$%
\begin{equation}
\mathcal{D}\varphi =E\varphi  \label{e1}
\end{equation}%
with $\mathcal{D}$ being a symmetric Schr\"{o}dinger operator of the form 
\begin{equation}
\mathcal{D}=-\sum_{k,l}\partial _{k}g_{kl}^{-1}\partial
_{l}+\sum_{k}r_{k}\partial _{k}~.  \label{D}
\end{equation}%
Here $g_{kl}^{-1}$ denotes the inverse of the curvature free symmetric
Riemannian tensor $g_{kl}=g_{lk}$, which, in view of (\ref{HJ}), is at most
quadratic in the coordinates $I_{i}$. The functions $r_{k}$ are assumed,
again in view of (\ref{HJ}), to be linear in the coordinates $I_{i}$. In
many cases these coordinates are taken to be invariant polynomials (for more
details see below), albeit sometimes re-parameterizations are needed to
guarantee the quadratic and linear dependence of $g_{kl}$ and $r_{k}$,
respectively.

Clearly, the operator (\ref{D}) is not of the usual form of a Hamiltonian,
that is Laplacian plus potential. In order to extract a potential from this
Hamiltonian one has to carry out a gauge transformation $\varphi =e^{\chi
}\psi $ to bring the equation (\ref{e1}) into the more standard form 
\begin{equation}
(-\Delta +V)\psi =E\psi  \label{e2}
\end{equation}%
involving the Laplace-Beltrami operator in general Riemannian space 
\begin{equation}
\Delta =\frac{1}{\sqrt{G}}\sum_{k,l}\partial _{k}\sqrt{G}g_{kl}^{-1}\partial
_{l}\qquad ~~~\text{with~~ }G^{-1}=\det g^{-1}
\end{equation}%
and a potential $V$. Extracting then from the equality $e^{-\chi }\mathcal{D}%
e^{\chi }=-\Delta +V$ the terms of first and zeroth order in $\partial _{l}$%
, one finds 
\begin{eqnarray}
r_{k} &=&\sum_{l}g_{kl}^{-1}\partial _{l}(2\chi -\ln \sqrt{G})  \label{r} \\
V &=&\sum_{k}r_{k}\partial _{k}\chi -\sum_{k,l}\left[ \partial
_{k}(g_{kl}^{-1}\partial _{l}\chi )+g_{kl}^{-1}\partial _{k}\chi \partial
_{l}\chi \right] ,  \label{V}
\end{eqnarray}%
respectively. Multiplying now (\ref{r}) with $g_{lk}^{-1}$ and
differentiating thereafter with $\partial _{m}$ one realizes that the right
hand side is symmetric under the exchange $m\leftrightarrow l$. Therefore,
one deduces immediately for the left hand side the same symmetry 
\begin{equation}
\partial _{m}\sum_{l}(g_{kl}r_{l})=\partial _{k}\sum_{l}(g_{ml}r_{l})~.
\label{re}
\end{equation}%
This equation constraints the values of $r_{k}$ and can be solved by 
\begin{equation}
r_{k}=\sum_{l}g_{kl}^{-1}\partial _{l}\rho ~.  \label{rr}
\end{equation}%
The function $\rho $ introduced at this point is referred to as
pre-potential. It should be stressed that there is no compelling argument in
this approach, which fixes this pre-potential and it remains subject to a
convenient ansatz. Substituting (\ref{rr}) back into (\ref{r}) and (\ref{V})
one then finds 
\begin{eqnarray}
\chi &=&\frac{1}{2}(\rho +\ln \sqrt{G})  \label{VV} \\
V &=&\frac{1}{4}\sum_{k,l}g_{kl}^{-1}\partial _{k}\rho \partial _{l}\rho -%
\frac{1}{4}\sum_{k,l}g_{kl}^{-1}\partial _{k}(\ln \sqrt{G})\partial _{l}(\ln 
\sqrt{G})-\sum_{k,l}\partial _{k}(g_{kl}^{-1}\partial _{l}\chi )~.
\label{vvv}
\end{eqnarray}%
It will turn out below that the term in $V$ which involves $\chi $ is zero
or constant. Provided that $g_{kl}^{-1}\partial _{l}(\ln \sqrt{G})$ is
linear in $\vec{I}$, this would follow directly as a consequence of the
assumption already made on $r_{k}$, namely that it is linear in the
variables $\vec{I}$. In that case we can deduce that this term in the
potential would be constant and omitting it just amounts to a constant shift
of the ground state energy.

Before we can specify in more detail the ansatz for the pre-potential $\rho $
proposed in \cite{HR3}, we have to gather various facts about invariant
polynomials. This will make the suggested ansatz look very natural, albeit
not entirely compelling.

\section{Polynomial invariants of the Coxeter group}

We specify here in more detail the nature of the variables $\vec{I}$ and
assemble some of their mathematical properties. First we recall the well
known fact, that to each simple root $\alpha _{i}$ in a root system $\Delta $
one can associate a reflection on the hyperplane through the origin
orthogonal to $\alpha _{i}$ 
\begin{equation}
\sigma _{i}(\vec{x})=\vec{x}-2\frac{\vec{x}\cdot \alpha _{i}}{\alpha _{i}^{2}%
}\alpha _{i}~\ \qquad \text{for }1\leq i\leq \ell ,~\vec{x}\in \mathbb{R}%
^{\ell }.  \label{def1}
\end{equation}%
These reflections constitute the Coxeter group $\mathcal{W}$ of rank $\ell $
or more specifically when $2\alpha \cdot \beta /\beta ^{2}\in \mathbb{Z}$
for all $\alpha ,\beta \in \Delta $ a Weyl group. One may then express each
vector $\vec{x}\in $ $\mathbb{R}^{\ell }$ as $\vec{x}=\sum_{i=1}^{\ell
}x_{i}\alpha _{i}$ and associate to it a polynomial $P(x_{1},\ldots ,x_{\ell
})$. The action of the Coxeter group on these polynomials is defined as 
\begin{equation}
\sigma _{i}P(x_{1},\ldots ,x_{\ell })=P(\sigma _{i}^{-1}(x_{1}),\ldots
,\sigma _{i}^{-1}(x_{\ell }))~.  \label{def2}
\end{equation}%
\noindent From the defining relations (\ref{def1}) and (\ref{def2}) it
follows directly that by taking the simple roots as a basis for $\mathbb{R}%
^{\ell }$ the action of the simple Weyl reflections acquires a particularly
simple form 
\begin{equation}
\sigma _{i}P(x_{1},...,x_{\ell
})=P(x_{1},...,x_{i-1},x_{i}-\sum_{j}x_{j}K_{ji},x_{i+1},...,x_{\ell })\;.
\label{sp}
\end{equation}%
Here $K$ denotes the Cartan matrix $K_{ij}=2\alpha _{i}\cdot \alpha
_{j}/\alpha _{j}^{2}$. The special set of polynomials which does not change
under the action of $\mathcal{W}$, i.e. for which 
\begin{equation}
\sigma _{i}I_{s}(x_{1},\ldots ,x_{\ell })=I_{s}(x_{1},\ldots ,x_{\ell
})~\qquad \text{for all }\sigma _{i}\in \mathcal{W}  \label{inv}
\end{equation}%
are the polynomial invariants of the Coxeter group. It turns out that a
basic set of linear independent polynomials $\left\{ I_{1+s_{1}},\ldots
,I_{1+s_{\ell }}\right\} $ can be graded by the $\ell $ exponents $s_{i}$ of
the Coxeter group, with $1\leq i\leq \ell $. The subscripts $1+s_{i}$
indicate here the degrees of the polynomials. It is this set of basic
invariants which one takes as the coordinates of the previously described
Hamiltonian system.

Let us now establish and recall some of their main properties, which we
shall exploit below:

\subsection{Eigenbasis of the Coxeter element}

It is clear that the choice of the basis for the coordinates will alter the
form of the potential and a priori there is no coordinate system which is
more special than another one. However, certain choices make the final
expressions very simple and one can take here the search for simplicity as a
guiding principle. A particularly suitable choice is the eigenbasis of the
Coxeter element. We will see that in this basis the expressions for the
polynomial invariants simplify considerably.

Adopting the notations of \cite{q2} (see also references therein), we first
define the Coxeter element $\sigma $ in terms of the two special elements of
the Weyl group 
\begin{equation}
\sigma _{\pm }:=\prod\limits_{\alpha _{i}\in \Phi _{\pm }}\sigma _{i}\,\,,
\label{spm}
\end{equation}%
as $\sigma :=\sigma _{-}\sigma _{+}$. Here we have partitioned the set of
simple roots into two disjoint sets, say $\alpha _{k}\in \Phi _{+}$ and $%
\beta _{k}\in \Phi _{-}$, by associating the values $c_{i}=\pm 1$ to the
vertices $i$ of the Dynkin diagram of the Lie algebra, in such a way that no
two vertices related to the same set are linked together. The eigensystem of
the Coxeter element can then be brought into the form 
\begin{equation}
\sigma v_{j}=e^{\frac{2\pi i}{h}s_{j}}v_{j}\qquad \text{and}\qquad
v_{j}=e^{-i\frac{\pi }{h}s_{j}}\sum_{k}\xi _{jk}\alpha _{k}+\sum_{k}\xi
_{jk}\beta _{k}\;,  \label{eigen}
\end{equation}%
where we denote by $\xi $ the matrix of left eigenvectors of the Cartan
matrix, i.e. 
\begin{equation}
\sum_{j=1}^{\ell }\xi _{ij}K_{jk}=4\sin ^{2}\frac{\pi s_{i}}{2h}\;\xi
_{ik}\;,  \label{eig}
\end{equation}%
and the $s_{i}$ are the aforementioned exponents. Then we implicitly define
a variable substitution $\{x_{i}\}\rightarrow \{w_{i}\}$ by the basis
transformation 
\begin{equation}
\vec{x}=\sum_{i}x_{i}\alpha _{i}=\sum_{i,j}\zeta _{ij}w_{j}\alpha
_{i}=\sum_{i}w_{i}v_{i}\;.  \label{xw}
\end{equation}%
with 
\begin{equation}
\zeta _{kj}:=\left\{ 
\begin{array}{cc}
e^{-i\frac{\pi }{h}s_{j}}\xi _{jk}\;, & \text{for \ }\alpha _{k}\in \Phi _{+}
\\ 
\xi _{jk}\;, & \text{for \ }\beta _{k}\in \Phi _{-}%
\end{array}%
\right. \;.  \label{xw2}
\end{equation}%
Defining then also polynomials in these new variables, we obtain as a
consequence of (\ref{eigen}) the action of the Coxeter element on these
polynomials 
\begin{equation}
\sigma P(w_{1},...,w_{\ell })=P(w_{1}e^{\frac{2\pi i}{h}},w_{2}e^{\frac{2\pi
i}{h}s_{2}},...,w_{\ell }e^{-\frac{2\pi i}{h}})\;.  \label{Coxact}
\end{equation}%
Recall here that $s_{i}+s_{\ell -i}=h$ \ for $1\leq i\leq \ell $. Since the
Coxeter element is build from simple Weyl reflections it follows from (\ref%
{inv}) that the invariants of the Coxeter group are also invariant under the
action of the Coxeter element 
\begin{equation}
\sigma I_{s}(x_{1},\ldots ,x_{\ell })=I_{s}(x_{1},\ldots ,x_{\ell
})~=I_{s}\left( \sum_{i}\zeta _{1i}w_{i},\ldots ,\sum_{i}\zeta _{\ell
i}w_{i}\right) ~.\qquad
\end{equation}%
To be able to compute the action of the Weyl reflections we have to express
the polynomials in terms of the $x$-variables. Nonetheless, by means of (\ref%
{xw}) we can also translate the action of the Weyl reflections in the $x$%
-variables to an action in terms of the $w$-variables which allows for a
more concise and possibly generic formulation of the invariants (see below).

\subsection{Universal formulae for invariants}

It is a natural question to ask, whether there exist formulae which express
the invariants in a universal fashion, that is valid for all algebras.
Indeed for invariants of degree 2 this is possible and we find 
\begin{equation}
I_{2}(\vec{x})=\sum_{i=1}^{\ell
}t_{i}x_{i}^{2}+\sum_{i<j}x_{i}K_{ij}t_{j}x_{j},\quad
K_{ij}t_{j}=K_{ji}t_{i},\quad t_{i}>1\;.
\end{equation}
Here the $t_{i}$ denote the symmetrizers of the Cartan matrix, which could
be avoided in the above expression by absorbing them into the roots, which
amounts to taking the simple co-roots instead of the simple roots as a
basis. For higher degrees we did not succeed to find universal formulae for
the invariants.

Changing, however, to the eigenbasis of the Coxeter element it is far more
obvious how to write down a universal expression. From (\ref{Coxact}) it is
clear that any invariant has to be of the form 
\begin{equation}
I_{s}(\vec{w})=\sum_{a_{1},\ldots ,a_{\ell }=1}^{s}c_{s}(a_{1},\ldots
,a_{\ell })w_{1}^{a_{1}}\ldots w_{\ell }^{a_{\ell }}  \label{gen}
\end{equation}
where the constants $c_{s}(a_{1},\ldots ,a_{\ell })$ are constrained as 
\begin{equation}
c_{s}(a_{1},\ldots ,a_{\ell })=\left\{ 
\begin{array}{l}
\neq 0\quad \text{if \ }\sum_{i=1}^{\ell }a_{i}=s,\quad \sum_{i=1}^{\ell
}a_{i}s_{i}=nh,\quad n\in \mathbb{Z} \\ 
=0\quad \text{otherwise}\qquad%
\end{array}
\right. ~.
\end{equation}
Consequently, this means for instance that the quadratic invariant has to be
of the form 
\begin{equation}
I_{2}=\sum\limits_{i}c_{2}(a_{i},a_{\ell -i+1})w_{i}w_{\ell -i+1}\;.
\end{equation}
One can proceed similarly for higher degrees, but it is then less obvious
how to fix the constants $c_{s}(a_{1},\ldots ,a_{\ell })$. Hence, for the
time being we have to rely on case-by-case studies, but even for explicit
algebras the most generic expressions are difficult to find in the
literature. See \cite{FK} for a complete list.

\subsection{Jacobians and Riemannians}

It will turn out that a key quantity in this scheme is the Jacobian
determinant related to the polynomial invariants 
\begin{equation}
J=\det \left( j\right) ~\qquad \text{with \ }j_{kl}=\frac{\partial
I_{1+s_{k}}}{\partial u_{l}}\text{\ }.  \label{Jac}
\end{equation}%
The determinant $J$ is known to possess various important properties, see
e.g. \cite{Hum2}:

\begin{description}
\item[i)] The polynomials $I_{1+s_{1}},\ldots ,I_{1+s_{\ell }}$ in $%
u_{1},\ldots ,u_{\ell }$ are algebraically independent if and only if $J\neq
0$.

\item[ii)] Defining for each root $\alpha $ a linear polynomial 
\begin{equation}
p_{\alpha }(u_{1},\ldots ,u_{\ell })=\sum_{i=1}^{\ell }\tilde{\gamma}%
_{\alpha }^{(i)}u_{i}~\qquad \qquad \text{with }\tilde{\gamma}_{\alpha
}^{(i)}\in \mathbb{R~},  \label{palph}
\end{equation}%
such that $p_{\alpha }(u_{1},\ldots ,u_{\ell })=0$ defines the hyperplane
through the origin orthogonal to $\alpha $. Then one can factorize $J$ as 
\begin{equation}
J=\mu \dprod\limits_{\alpha \in \Delta _{+}}p_{\alpha }(u_{1},\ldots
,u_{\ell })~\qquad \qquad \text{with }\mu \in \mathbb{R~},  \label{J2}
\end{equation}%
where $\Delta _{+}$ denotes the set of positive roots.

\item[iii)] 

Defining the inverse Riemannian in terms of the basic invariants $\left\{
I_{1+s_{1}},\ldots ,I_{1+s_{\ell }}\right\} $%
\begin{equation}
g_{kl}^{-1}=\sum_{i=1}^{\ell }\frac{\partial I_{k}}{\partial u_{i}}\frac{%
\partial I_{l}}{\partial u_{i}}~,  \label{invg}
\end{equation}%
the Jacobian determinant is related to the determinant of the inverse
Riemannian as 
\begin{equation}
J^{2}=G^{-1}=\det g^{-1}~.  \label{JG}
\end{equation}
\end{description}

\noindent The factorization properties (\ref{J2}) and (\ref{JG}) for $J$ and 
$G^{-1}$, respectively, will make the ansatz for the pre-potential $\rho $
appear very natural.

It is worth pointing out that the choice of the metric (\ref{invg})
guarantees that the Laplacian in the variables $\vec{u}$ is flat. This is
easily seen by considering the change of the Euclidean metric tensor, i.e. $%
g(\vec{u})_{mn}=\delta _{mn}$, under a coordinate transformation. For this
we just have to multiply 
\begin{equation}
g(I)_{ij}=\sum_{m,n}\frac{\partial u^{m}}{\partial I_{i}}\frac{\partial u^{n}%
}{\partial I_{j}}\,g(u)_{mn}=\sum_{m}\frac{\partial u^{m}}{\partial I_{i}}%
\frac{\partial u^{m}}{\partial I_{j}},\quad 
\end{equation}%
with (\ref{invg}). This choice avoids the entire analysis which is needed in
this approach to guarantee the flatness of the Laplacian as carried out in 
\cite{HR1}.

\section{Polynomial invariants of the q-deformed Coxeter group}

We extend now the previous discussion and seek polynomials which are
invariant under $q$-deformed Weyl reflections. We adopt here the notation of 
\cite{q2}, (see also \cite{q1}), for more details on $q$-deformed Weyl
reflections and the general context in which they emerged. When acting on a
simple root they are defined as 
\begin{equation}
\sigma _{i}^{q}(\alpha _{j})=\alpha _{j}-(2\delta _{ij}-[I_{ji}]_{q})\alpha
_{i}  \label{qd}
\end{equation}
with $I=2-K$ \ being the incidence matrix of some Lie algebra and $%
[n]_{q}=(q^{n}-q^{-n})/(q-q^{-1})$ being the standard notation for a $q$%
-deformed integer. According to the notions outlined in the previous section
the invariants are obviously defined by 
\begin{equation}
\sigma _{i}^{q}I_{s}(x_{1},\ldots ,x_{\ell })=I_{s}(x_{1},\ldots ,x_{\ell
})~\qquad \text{for }1\leq i\leq \ell ~.  \label{inq}
\end{equation}

\noindent Once again for polynomials of degree 2 we can write down a
universal formula 
\begin{equation}
I_{2}^{q}=\sum_{i=1}^{\ell
}[t_{i}]_{q}x_{i}^{2}+\sum_{i<j}x_{i}[K_{ij}]_{q}[t_{j}]_{q}x_{j}  \label{i2}
\end{equation}
which is invariant under the $q$-deformed Weyl reflections. As in the
non-deformed case higher invariants are difficult to write down in a
universal form. Note also that the $q$-deformed invariants in the sense of (%
\ref{inq}) are not invariants of the $q$-deformed Coxeter elements as
defined in \cite{q1,q2}, as the latter does not just consist of
transformations $\sigma _{i}^{q}$. However, we can alter this definition
slightly to achieve this, see section 7.

\section{Ansatz for the pre-potential}

Having fixed a set of basic invariants $\left\{ I_{1+s_{1}},\ldots
,I_{1+s_{\ell }}\right\} $ one assumes that the wavefunctions in (\ref{e1})
and (\ref{e2}) depend on these coordinates, that is $\varphi \rightarrow
\varphi (\vec{I})$ and $\psi \rightarrow \psi (\vec{I})$. Naturally one can
then also view the potential as a function of the invariants, i.e. $%
V\rightarrow V(\vec{I})$ and understand that $\partial _{k}\equiv \partial
/\partial I_{k}$. Defining the inverse Riemannian as in (\ref{invg}) and
using (\ref{JG}), we can re-write the potential (\ref{vvv}) in the form 
\begin{equation}
V=\frac{1}{4}\sum_{k,l}g_{kl}^{-1}\partial _{k}\rho \partial _{l}\rho -\frac{%
1}{4J^{2}}\sum_{k,l}g_{kl}^{-1}\partial _{k}J\partial _{l}J~-\frac{1}{2}%
\sum_{k,l}\partial _{k}\left[ g_{kl}^{-1}~\left( \partial _{l}\rho -\frac{1}{%
J}\partial _{l}J\right) \right] .  \label{JP}
\end{equation}

\noindent \noindent For the above mentioned reason the last term will
usually drop out. It is apparent from this formulation, that re-scaling $J$
by a constant will not alter the potential, a fact which is important with
regard to the occurrence of possible coupling constants. To be more specific
about the potential one has to choose a suitable pre-potential. In \cite%
{HR2,HR3,HR4} the following ansatz was proposed 
\begin{equation}
\rho =\dsum\limits_{i=0}^{\ell }\gamma _{i}\ln P_{i}(\vec{I}).  \label{anpre}
\end{equation}%
The $P_{i}(\vec{I})$ for $1\leq i\leq \ell $ are defined by the
factorization of the determinant of the inverse Riemannian 
\begin{equation}
J^{2}=G^{-1}=\det g^{-1}=\prod_{i=1}^{\ell }P_{i}(\vec{I})~.  \label{pre}
\end{equation}%
Evidently, this ansatz (\ref{pre}) is inspired by the properties ii) and
iii). However, there is an additional significant constraint namely that the 
$P_{i}(\vec{I})$ are functions of the invariants $\vec{I}$, which as was
argued above is needed to guarantee the solvability. Often one would like to
obtain also an additional harmonic confining term proportional to $%
\dsum_{i}u_{i}^{2}$ in $V$. This is easily achieved by including also a
factor of the form $P_{0}\sim \exp (\dsum_{i}u_{i}^{2})$ into the ansatz for
the pre-potential (\ref{anpre}). For the reasons outlined in section 2, the
entire Hamiltonian, that means also this term, has to be expressed in terms
of invariant polynomials. Usually we can take $P_{0}=\exp (I_{2})$.

\noindent Substituting the ansatz (\ref{pre}) into (\ref{vvv}), the
potential acquires the form 
\begin{equation}
V=\frac{1}{4}\sum_{i,j,k,l}\left( \gamma _{i}\gamma _{j}-\frac{1}{4}\right)
g_{kl}^{-1}\partial _{k}(\ln P_{i})\partial _{l}(\ln P_{j})-\frac{1}{2}%
\sum_{i,k,l}\left( \gamma _{i}-\frac{1}{2}\right) \partial _{k}\left[
g_{kl}^{-1}\partial _{l}(\ln P_{i})\right] ~.  \label{pot}
\end{equation}%
As was pointed out in \cite{HR3}, it will turn out that the terms with $%
i\neq j$ in the first term are constants (even zero) most of the time, which
therefore can be dropped safely by just shifting the ground state. The other
motivation for the ansatz (\ref{pre}) is that one would like the two terms
in (\ref{JP}) or (\ref{vvv}) to combine naturally.

Comparing now (\ref{pre}), (\ref{J2}) and (\ref{JG}) we can proceed and
exploit the fact that the $P_{i}$ factorize further into linear polynomials 
\begin{equation}
P_{i}=\dprod\limits_{\alpha \in \Delta _{+}^{(i)}}\left[ p_{\alpha }(\vec{u})%
\right] ^{2}~.  \label{ff}
\end{equation}%
The relation (\ref{ff}) is here the defining relation for the set of
positive roots $\Delta _{+}^{(i)}$. Exploiting the fact that the $p_{\alpha
}(\vec{u})$ are linear in $\vec{u}$, see (\ref{palph}), and changing from
the invariant polynomials as coordinates to the $\vec{u}$-variables, we
obtain 
\begin{equation}
V=\sum_{i,j}\left[ \left( \gamma _{i}\gamma _{j}-\frac{1}{4}\right)
\sum_{\alpha \in \Delta _{+}^{(i)},\beta \in \Delta _{+}^{(j)}}\frac{\sum_{k}%
\tilde{\gamma}_{\alpha }^{(k)}\tilde{\gamma}_{\beta }^{(k)}}{p_{\alpha }(%
\vec{u})p_{\beta }(\vec{u})}\right] ~.  \label{pott}
\end{equation}%
Recall that the $\tilde{\gamma}_{\alpha }^{(k)}$ are defined in (\ref{palph}%
). The expression (\ref{pott}) constitutes a general formula for potentials
when starting with \textit{any }Coxeter group. This structure will survive
in different coordinate systems, as clearly the $p_{\alpha }(\vec{u})$
remain linear after linear coordinate transformations. Often one changes the
coordinate system by using an explicit representations for the roots in an
orthogonal basis, which one may find in various places of the literature,
e.g. \cite{Hum2}. Then (\ref{pott}) enables one to write down directly the
potentials associated to any Coxeter group. In practice it turns out that
only the diagonal terms in the sum survive, i.e. $\alpha =\beta $, such that
the potentials will always be of Calogero type. We omitted in (\ref{pott})
the term resulting from the last term in (\ref{pot}) and also a possible
constant.

\section{From G$_{2}$-polynomial invariants to the Calogero models}

In \cite{Cap} it was shown that the Calogero models for three particles
exhibit an underlying $G_{2}$-structure, which can be exploited to establish
their solvability. In \cite{HR3} this procedure was reversed and it was
shown that the approach outlined in section 2 and 3 indeed yields potentials
of the Calogero type when one starts with a $G_{2}$-structure. Here we
recall briefly the procedure, mainly to set the scene for the q-deformed
treatment below, but also to establish a few facts not pointed out so far.
In particular, in \cite{Cap} as well as in \cite{HR3} not the most general $%
G_{2}$-invariants were used. As we will show below the most generic
invariants involve some arbitrary constants. The obvious question to ask is
whether one obtains a new type of potential when using the procedure
outlined above in terms of these generic coordinates, possibly involving
additional coupling constants.

Let us start with the action of the $G_{2}$-Weyl reflections on the simple
roots 
\begin{eqnarray}
\sigma _{1}(\alpha _{1}) &=&-\alpha _{1}\qquad \qquad ~~~~\sigma _{1}(\alpha
_{2})=3\alpha _{1}+\alpha _{2} \\
\sigma _{2}(\alpha _{1}) &=&\alpha _{1}+\alpha _{2}\qquad \qquad \sigma
_{2}(\alpha _{2})=-\alpha _{2}~.
\end{eqnarray}%
Then for a general vector $\vec{x}=x_{1}\alpha _{1}+x_{2}\alpha _{2}$ in $%
\mathbb{R}^{2}$ we have 
\begin{eqnarray}
\sigma _{1}(\vec{x}) &=&(3x_{2}-x_{1})\alpha _{1}+x_{2}\alpha _{2} \\
\sigma _{2}(\vec{x}) &=&x_{1}\alpha _{1}+(x_{1}-x_{2})\alpha _{2},
\end{eqnarray}%
such that 
\begin{eqnarray}
\sigma _{1}P(x_{1},x_{2}) &=&P(3x_{2}-x_{1},x_{2})  \label{s1} \\
\sigma _{2}P(x_{1},x_{2}) &=&P(x_{1},x_{1}-x_{2})~.  \label{s2}
\end{eqnarray}%
The equations (\ref{s1}) and (\ref{s2}) follow also directly from (\ref{sp}).

Using (\ref{s1}) and (\ref{s2}) we can now generate solutions to the
equation (\ref{inv}), i.e. construct the invariant polynomials. The
procedure is straightforward. We simply write down the most generic
expression for a potential candidate for a polynomial invariant $I_{s}(\vec{x%
})$ of degree $s$ similar to the form as in (\ref{gen}) with arbitrary
coefficients $c_{s}$, but now in terms of the $x$-variables. Acting then
with all simple Weyl reflections on this polynomial and demanding invariance
(\ref{inv}) leads to a system of equations which determine the $c_{s}$.
Depending on the degree and the algebra this might not yield enough
equations to fix all constants and one ends up with expressions still
involving free parameters. In this manner we find as generic invariants 
\begin{eqnarray}
I_{2} &=&\kappa _{2}\left( \frac{1}{3}x_{1}^{2}+x_{2}^{2}-x_{1}x_{2}\right)
\label{G2I2} \\
I_{6} &=&\kappa _{6}\left( -\frac{2}{27}x_{1}^{6}+\frac{2}{3}x_{1}^{5}x_{2}-%
\frac{5}{3}x_{1}^{4}x_{2}^{2}+5x_{1}^{2}x_{2}^{4}-6x_{1}x_{2}^{5}+2x_{2}^{6}%
\right) +\tilde{\kappa}_{6}I_{2}^{3}~.  \label{G2I6}
\end{eqnarray}%
Here $\kappa _{2}$, $\kappa _{6}$ and $\tilde{\kappa}_{6}$ remain arbitrary
constants. We see that besides an overall constant, which is naturally
always present, $I_{6}$ also involves an additional free parameter $\tilde{%
\kappa}_{6}$.

A first restriction on possible values the constants might take comes from
the fact that we want $I_{2}$ and $I_{6}$ to be algebraically independent.
To establish this we compute first the Jacobian determinant for these
invariants according to the definition (\ref{Jac})

\begin{equation}
J=\frac{2}{3}\kappa _{2}\kappa
_{6}x_{1}x_{2}(x_{1}-x_{2})(x_{1}-2x_{2})(x_{1}-3x_{2})(2x_{1}-3x_{2})~.
\label{factJ}
\end{equation}
As we expect from property i) stated in section 3 and the explicit
expression in (\ref{G2I6}) we have to keep $\kappa _{6}\neq 0$ in order to
guarantee the algebraic independence of $I_{2}$ and $I_{6}$. Obviously, for $%
\kappa _{6}=0$ we have $I_{6}=\tilde{\kappa}_{6}I_{2}^{3}$. We also note
that $\tilde{\kappa}_{6}$ can remain completely arbitrary in this context,
but as we see below, we can not simply set it to zero for our purposes.

Alternatively, we can compute the Jacobian determinant by an entirely
different formula, namely (\ref{J2}), and thus confirm the computation which
led to (\ref{factJ}). To be able to use (\ref{J2}), we recall that the
positive roots of $G_{2}$ are, (see e.g. \cite{Hum,Hum2})) 
\begin{equation}
\Delta _{+}^{G_{2}}=\{\alpha _{1},\alpha _{2},\alpha _{1}+\alpha
_{2},2\alpha _{1}+\alpha _{2},3\alpha _{1}+\alpha _{2},3\alpha _{1}+2\alpha
_{2}\}~,
\end{equation}%
such that the hyperplanes result to 
\begin{equation}
\begin{array}{lll}
p_{\alpha _{1}}=2x_{1}-3x_{2}, & ~p_{\alpha _{1}+\alpha
_{2}}=x_{1}-3x_{2},~~~~ & p_{2\alpha _{1}+\alpha _{2}}=x_{1}, \\ 
p_{3\alpha _{1}+\alpha _{2}}=x_{1}-x_{2},~~~~ & p_{\alpha _{2}}=x_{1}-2x_{2},
& ~p_{3\alpha _{1}+2\alpha _{2}}=x_{1}.%
\end{array}
\label{h1}
\end{equation}%
Assembling the $p_{\alpha }$ into the product (\ref{J2}) the result confirms
our findings (\ref{factJ}) with $\mu =(2/3)\kappa _{2}\kappa _{6}$. Notice,
that the constants $\kappa _{2}$ and $\kappa _{6}$ organize in a separate
factor in $J$, such that different choices, apart from $\kappa _{2}=0$ or $%
\kappa _{6}=0$, will not alter the polynomial structure of $J$. From (\ref%
{JP}) we deduce that in the second term of the potential the overall factor
in $J$ just cancels, such that the constants $\kappa _{i}$ completely drop
out from this term.

So far, we have only two coordinates. To incorporate a three body
interaction we need one more coordinate. Let us therefore choose an
orthogonal basis in $\mathbb{R}^{3}$ for the simple roots $\alpha
_{1}=\varepsilon _{1}-$ $\varepsilon _{2}$ and $\alpha _{2}=-2\varepsilon
_{1}+$ $\varepsilon _{2}+$ $\varepsilon _{3}$, with $\varepsilon _{i}\cdot
\varepsilon _{j}=\delta _{ij}$ (see e.g. \cite{Hum2}). Then we can introduce
a new set of variables via the relation 
\begin{equation}
\vec{x}=x_{1}\alpha _{1}+x_{2}\alpha _{2}=(x_{1}-2x_{2})\varepsilon
_{1}+(x_{2}-x_{1})\varepsilon _{2}+x_{2}\varepsilon _{3}=y_{1}\varepsilon
_{1}+y_{2}\varepsilon _{2}+y_{3}\varepsilon _{3}~,
\end{equation}%
with the built-in constraint $y_{1}+y_{2}+y_{3}=0$. In these variables the
action of the Weyl group on the same polynomials becomes 
\begin{eqnarray}
\sigma _{1}P(y_{1},y_{2},y_{3}) &=&P(y_{2},y_{1},y_{3})  \label{y1} \\
\sigma _{2}P(y_{1},y_{2},y_{3}) &=&P(-y_{1},-y_{3},-y_{2})~.  \label{y2}
\end{eqnarray}%
In principle, we could proceed as above, i.e. writing down generic
expression with arbitrary coefficients and use directly (\ref{y1}) and (\ref%
{y2}) to find invariant polynomials in the $y$-variables. However, since we
have increased the number of coordinates also the amount of unknown
coefficients grows and we will end up with polynomials involving many more
free constants than just the three we are left with when using the $x$%
-variables. Instead, as we know the invariants already, we can also use in (%
\ref{G2I2}) and (\ref{G2I6}) directly the substitutions $x_{1}\rightarrow
-y_{1}-2y_{2}$ and $x_{2}\rightarrow -y_{1}-y_{2}$, such that we obtain 
\begin{eqnarray}
I_{2} &=&\kappa _{2}(y_{2}^{2}+y_{1}^{2}+y_{1}y_{2})/3  \label{Iy1} \\
I_{6} &=&\frac{1}{27}(\tilde{\kappa}_{6}-2\kappa _{6})(y_{1}^{6}+y_{2}^{6})+%
\frac{1}{9}(\tilde{\kappa}_{6}-2\kappa _{6})(y_{1}^{5}y_{2}+y_{1}y_{2}^{5}) 
\notag \\
&&+\frac{1}{9}(5\kappa _{6}+2\tilde{\kappa}%
_{6})(y_{1}^{4}y_{2}^{2}+y_{1}^{2}y_{2}^{4})+\frac{1}{27}(40\kappa _{6}+7%
\tilde{\kappa}_{6})y_{1}^{3}y_{2}^{3}.~~  \label{Iy3}
\end{eqnarray}%
Only for the special choice $\kappa _{2}=-3$, $\tilde{\kappa}_{6}=2\kappa
_{6}$ and $\kappa _{6}=1$ the expressions (\ref{Iy1}) and (\ref{Iy3}) reduce
to the coordinates $\lambda _{1}$, $\lambda _{2}$ used in \cite{Cap}.
Notice, that in the $y$-variables the invariants become symmetric
polynomials \cite{Symm}.

In order to reproduce the Calogero potentials we need to make yet another
coordinate transformation and introduce Jacobi relative coordinates 
\begin{equation}
y_{i}=z_{i}-\frac{1}{3}\sum_{j=1}^{3}z_{j}~,  \label{JR}
\end{equation}

\noindent which separate off the center of mass motion. The constraint in
the $y$-coordinates is now replaced by $z_{1}+z_{2}+z_{3}=3Z$, where $Z$ is
constant. In these coordinates we compute the inverse Riemannian (\ref{invg}%
) to 
\begin{equation}
g_{ij}^{-1}=\sum_{k=1}^{3}\frac{\partial I_{i}}{\partial z_{k}}\frac{%
\partial I_{j}}{\partial z_{k}}=\kappa _{2}\left( 
\begin{array}{cc}
2/3I_{2} & 2I_{6} \\ 
2I_{6} & ~~6I_{2}^{2}/\kappa _{2}^{3}~\left[ 2\tilde{\kappa}%
_{6}I_{6}+I_{2}^{3}(4\kappa _{6}^{2}-\tilde{\kappa}_{6}^{2})/\kappa _{2}^{3}%
\right]%
\end{array}%
\right) _{ij}~.  \label{g}
\end{equation}%
Apparently $g_{ij}^{-1}$ is not quadratic in the variables $I_{i}$, which as
we discussed is a necessary requirement to be able to bring the Hamiltonian
into the form (\ref{HJ}) and hence ensuring the solvability of the model.
However, for $\tilde{\kappa}_{6}=2\kappa _{6}$ one can choose a different
set of variables $I_{2}=\tau _{2}$, $I_{6}=\tau _{3}^{2}$, see \cite{Cap},
such that $\partial _{i}g_{ij}^{-1}\partial _{j}$ is of the desired form in
the $\tau $ variables. Alternatively, one can also use the representation of
a different algebra to establish solvability \cite{Cap}. Let us from now on
take $\tilde{\kappa}_{6}=2\kappa _{6}$ and $\kappa _{2}=1$ but leaving $%
\kappa _{6}$ arbitrary. Then we compute from (\ref{g}) 
\begin{equation}
G^{-1}=\det g^{-1}=4I_{6}\left( 4\kappa _{6}I_{2}^{3}-I_{6}\right) ~.
\end{equation}%
\noindent For the pre-potential we make now an ansatz according to (\ref{pre}%
), where we also include the previously mentioned $P_{0}$-term 
\begin{equation}
P_{0}=e^{I_{2}},\qquad P_{1}=4\kappa _{6}I_{2}^{3}-I_{6},\qquad P_{2}=I_{6}~.
\end{equation}%
\noindent \noindent From this we compute with formula (\ref{pot}) the
potential to

\begin{eqnarray}
V &=&\frac{\gamma _{0}^{2}}{6}I_{2}+3\lambda _{1}\kappa _{6}\frac{I_{2}^{2}}{%
4\kappa _{6}I_{2}^{3}-I_{6}}+3\lambda _{2}\kappa _{6}\frac{I_{2}^{2}}{I_{6}}
\label{v1} \\
&=&\frac{1}{2}\omega ^{2}\dsum\limits_{k=1}^{3}z_{k}^{2}+\lambda
_{1}\dsum\limits_{1\leq i<j\leq 3}\frac{1}{(z_{i}-z_{j})^{2}}+3\lambda
_{2}\dsum\limits_{\substack{ 1\leq i<j\leq 3  \\ i,j\neq k}}\frac{1}{%
(z_{i}+z_{j}-2z_{k})^{2}}~,  \label{v2}
\end{eqnarray}%
where the coupling constants are $\omega =\gamma _{0}/(3\sqrt{2})$ and $%
\lambda _{i}=2\gamma _{i}^{2}-1/2$ for $i=1,2$. The potential in the form (%
\ref{v2}) corresponds to so-called rational $G_{2}$-model \cite{Wolf}, which
reduces to the Calogero model \cite{Cal1,Cal2,Cal3} when the three-particle
interaction is switched off, i.e. for $\lambda _{2}\rightarrow 0$. Notice
that the coupling constants \ $\gamma _{0}$, $\gamma _{1}$ and $\gamma _{2}$
which enter the scheme through the ansatz for the pre-potential just
reparameterize the coupling constants of the $G_{2}$-model. Note also that
the constant $\kappa _{6}$ has dropped out completely, such that any choice,
apart from $\kappa _{6}=0$, will yield the same potential (\ref{v2}). Hence,
the ambiguity in the choice of the invariant polynomials as coordinates has
no bearing on the physics.

The invariants acquire a particularly simple form when we use the eigenbasis
of the Coxeter element. The transformations outlined above yield for the $%
G_{2}$-case 
\begin{equation}
x_{1}=\sqrt{3}(e^{-i\pi /6}w_{1}-e^{-i\pi 5/6}w_{2}),\quad
x_{2}=w_{1}+w_{2}\;,
\end{equation}%
such that the invariants simplify considerably 
\begin{equation}
I_{2}=\kappa _{2}w_{1}w_{2}\quad \text{and\quad }I_{6}=\kappa
_{6}(w_{1}^{6}+w_{2}^{6})+\tilde{\kappa}_{6}w_{1}^{3}w_{2}^{3}~.
\end{equation}%
We then find for the Jacobian 
\begin{eqnarray}
J &=&-6\kappa _{2}\kappa _{6}(w_{1}^{6}-w_{2}^{6}) \\
&=&-6\kappa _{2}\kappa
_{6}(w_{1}+w_{2})(w_{1}-w_{2})(w_{1}^{2}-w_{1}w_{2}+w_{2}^{2})(w_{1}^{2}+w_{1}w_{2}+w_{2}^{2})~,
\end{eqnarray}%
which could of course be used to construct the potential in these variables.
As this type of \ factorization of $J$ involves quadratic polynomials we
will end up with potentials not quite of Calogero type. To achieve this we
would have to factorize the last two terms further involving complex
coefficients, but in that case the individual two particle interactions
terms would be complex. Solvability is only guaranteed when we can express
the factors in terms of the invariant polynomials.

\section{Exactly solvable potentials from q-deformed G$_{2}$-polynomial
invariants}

Let us now extend the previous analysis to the q-deformed case. To commence
we need to evaluate the q-deformed Weyl reflections $\sigma _{i}^{q}$ as
defined in (\ref{qd}) for which we require the q-deformed Cartan matrix. In
our conventions it reads for the $G_{2}$-case 
\begin{equation}
K_{q}=\left( 
\begin{array}{cc}
2 & -1 \\ 
-[3]_{q} & 2%
\end{array}%
\right) ~.
\end{equation}%
With $K_{q}$ at hand we can now seek invariant polynomials according to the
definitions (\ref{qd}) and (\ref{inq}). We proceed in the same manner as for
the non-deformed case and start with generic expressions $I_{s}^{q}(\vec{x})$
for polynomials of degree $s$ as in (\ref{gen}) and fix the constants as
outlined above. For generic deformation parameters $q$ we did not find
invariants. However, if we parameterize the $q$'s as 
\begin{equation}
q^{2}=\frac{1}{2}\left( 1+2\cos \frac{2\pi }{h}\right) +\sqrt{\left( 1+2\cos 
\frac{2\pi }{h}\right) ^{2}-4}  \label{qq}
\end{equation}%
with $h$ being some integer, the q-deformed Cartan matrix becomes 
\begin{equation}
(K_{q})_{ij}=\frac{2\alpha _{i}^{q}\cdot \alpha _{j}^{q}}{\alpha
_{j}^{q}\cdot \alpha _{j}^{q}}=\left( 
\begin{array}{cc}
2 & -1 \\ 
-4\cos ^{2}\frac{\pi }{h} & 2%
\end{array}%
\right) _{ij}~.  \label{kq}
\end{equation}%
Implicitly, we used here (\ref{kq}) to define some q-deformed roots $\alpha
_{i}^{q}$. Clearly for $h=3$ we recover the Cartan matrix of $A_{2}$, for $%
h=4$ we obtain the one of $C_{2}$ and $h=6$ corresponds to $G_{2}$. For the
values (\ref{qq}) of $q$ we find there exist always the invariants $%
I_{2}^{q} $ and $I_{h}^{q}$. From the above mentioned arguments this
suggests that the exponents of this algebra $G_{2}^{q}$ are $1$ and $h-1$.
This assertion is supported by the observation that the formula for
eigenvalues of the Cartan matrix (\ref{eig}) still holds for the q-deformed
case (\ref{kq}) when taking $s_{1}=1$ and $s_{2}=h-1$.

\noindent As we argued in the previous section, it is difficult to find
generic expressions for the invariants in the $x$-variables. However, as we
will see in the eigenbasis of the Coxeter element this task simplifies
drastically. According to (\ref{xw}) and (\ref{xw2}) we have the
transformations 
\begin{equation}
\vec{x}=\left( 
\begin{array}{cc}
1+e^{-\frac{2\pi i}{h}} & 1+e^{\frac{2\pi i}{h}} \\ 
1 & 1%
\end{array}%
\right) \vec{w}\quad ~~\Leftrightarrow ~~\quad \vec{w}=\frac{i}{2\sin \frac{%
2\pi }{h}}\left( 
\begin{array}{rr}
1 & -1-e^{\frac{2\pi i}{h}} \\ 
-1 & 1+e^{-\frac{2\pi i}{h}}%
\end{array}%
\right) \vec{x}~,  \label{trans}
\end{equation}%
such that the q-deformed Weyl reflections in the $w$-variables simplify to 
\begin{equation}
\sigma _{1}^{q}(w_{1})=w_{2},\quad \sigma _{1}^{q}(w_{2})=w_{1},\quad \sigma
_{2}^{q}(w_{1})=e^{\frac{2\pi i}{h}}w_{2},\quad \sigma _{2}^{q}(w_{2})=e^{-%
\frac{2\pi i}{h}}w_{1}~.  \label{sw}
\end{equation}%
The $\sigma _{1}^{q}$-transformations dictate that the invariants have to be
symmetric in $w_{1}$, $w_{2}$ and the $\sigma _{2}^{q}$-transformations
constrain their overall degree. With (\ref{sw}) we can easily find the most
generic expressions for the invariants 
\begin{equation}
I_{2}^{q}=\kappa _{2}w_{1}w_{2}\text{\qquad and\qquad }I_{h}^{q}=\kappa
_{h}(w_{1}^{h}+w_{2}^{h})~+\tilde{\kappa}_{h}(w_{1}w_{2})^{h/2}.  \label{Iq}
\end{equation}%
where $\tilde{\kappa}_{h}=0$ for $h$ being an odd integer. We can now
transform back to the $x$-variables and confirm for instance the generic
formula (\ref{i2}) for the invariant of degree 2, which still takes on a
fairly simple form 
\begin{equation}
I_{2}^{q}=\frac{\kappa _{2}}{4\sin ^{2}(2\pi /h)}\left( x_{1}^{2}+4\cos ^{2}%
\frac{\pi }{h}x_{2}^{2}-4\cos ^{2}\frac{\pi }{h}x_{1}x_{2}\right) ~.
\label{Iq2}
\end{equation}%
On the other hand, the expressions for the $I_{h}^{q}$ are already quite
cumbersome, albeit it is clear how to construct them from (\ref{Iq}) and (%
\ref{trans}).

As we saw in the previous section it was crucial to change the coordinate
system yet further to recover the Calogero potentials in the usual form. We
proceed here similarly. Let us choose first an orthogonal basis for the two
simple q-deformed roots in $\mathbb{R}^{3}$%
\begin{eqnarray}
\alpha _{1}^{q} &=&(\sqrt{3}\cos \frac{2\pi }{h}+\sin \frac{2\pi }{h},-2\sin 
\frac{2\pi }{h},\sin \frac{2\pi }{h}-\sqrt{3}\cos \frac{2\pi }{h})/\sqrt{3}
\label{o1} \\
\alpha _{2}^{q} &=&(-\sqrt{3}-\sqrt{3}\cos \frac{2\pi }{h}-\sin \frac{2\pi }{%
h},2\sin \frac{2\pi }{h},\sqrt{3}+\sqrt{3}\cos \frac{2\pi }{h}-\sin \frac{%
2\pi }{h})/\sqrt{3}~.  \label{o2}
\end{eqnarray}%
The inner products of these roots are $\alpha _{1}^{q}\cdot \alpha
_{1}^{q}=2 $, $\alpha _{2}^{q}\cdot \alpha _{2}^{q}=8\cos ^{2}\frac{\pi }{h}$
and $\alpha _{1}^{q}\cdot \alpha _{2}^{q}=-4\cos ^{2}\frac{\pi }{h}$ such
that we recover the q-deformed Cartan matrix according to (\ref{kq}). Of
course the choices (\ref{o1}) and (\ref{o2}) are not uniquely determined. As
an additional selection criterion we demand that $I_{2}^{q}$ will be of an
analogous form (\ref{Iy1}) as in the non-deformed case for all choices of $h$%
, such that it will be ensured that we can express $g_{22}^{-1}$ in terms of 
$I_{2}^{q}$. At the same time this will ensure that $P_{0}=\exp (I_{2}^{q})$
yields the harmonic confining potential, similarly as for the standard $%
G_{2} $-case. The above choice for the simple roots induces a definition for
new variables 
\begin{equation}
\vec{y}=x_{1}\alpha _{1}^{q}+x_{2}\alpha _{2}^{q}~,
\end{equation}%
which satisfy the constraint $y_{1}+y_{2}+y_{3}=0$. In turn this means that
we can replace in (\ref{Iq2}) 
\begin{equation}
\vec{x}=-\frac{1}{2}\left( 
\begin{array}{cc}
2 & ~~~(1+\sqrt{3}\cot \frac{\pi }{h}) \\ 
2 & ~~\ (1+\sqrt{3}\cot \frac{2\pi }{h})%
\end{array}%
\right) \vec{y}
\end{equation}%
such that $I_{2}^{q}$ is indeed always of the form (\ref{Iy1}) 
\begin{equation}
I_{2}^{q}=\frac{\kappa _{2}}{4\sin ^{2}(2\pi /h)}\left(
y_{1}^{2}+y_{2}^{2}+y_{1}y_{2}\right) =\frac{3\kappa _{2}}{8\sin ^{2}(2\pi
/h)}\left( \sum\limits_{i=1}^{3}z_{i}^{2}-3Z^{2}\right) ~.
\end{equation}%
Again it is obvious how to obtain the expressions for $I_{h}^{q}$, but they
turn out to be more cumbersome. From (\ref{Iq}) it is apparent that the case
of even and odd Coxeter number exhibit different behaviour. We treat them
now separately and supply for each case an explicit example.

\subsection{Even Coxeter numbers, h=8}

As the obvious difference between the odd and even case we found that in the
even case the additional constant $\tilde{\kappa}_{h}$ is entering the
procedure. We shall see that there are also other more profound differences.
We present the case $h=8$ in detail. For this the relation (\ref{qq}) simply
yields 
\begin{equation}
q^{2}=\frac{1}{2}(1+\sqrt{2})+\sqrt{2\sqrt{2}-1},\qquad \text{and\qquad }%
[3]_{q}=2+\sqrt{2}.
\end{equation}%
We can now take the invariants as given by the relations (\ref{Iq}) and
carry out the substitutions $\vec{w}\rightarrow \vec{x}\rightarrow \vec{y}%
\rightarrow \vec{z}$ specified above. In terms of the Jacobian relative
coordinates $z_{i}$ the inverse Riemannian (\ref{invg}) results to 
\begin{equation}
g_{ij}^{-1}=\sum_{k=1}^{3}\frac{\partial I_{i}}{\partial z_{k}}\frac{%
\partial I_{j}}{\partial z_{k}}=\kappa _{2}\left( 
\begin{array}{cc}
I_{2} & 4I_{8} \\ 
4I_{8} & ~~~~~16I_{2}^{3}/\kappa _{2}^{4}~\left[ 2\tilde{\kappa}%
_{8}I_{8}+I_{2}^{4}(\tilde{\kappa}_{8}^{2}-4\kappa _{8}^{2})/\kappa _{2}^{4}%
\right]%
\end{array}%
\right) _{ij}~.
\end{equation}

At this point we are facing a similar problem as in the non-deformed case,
that is $h=6$, namely that $g_{ij}^{-1}$ is not of degree two in the
variables $I_{i}$. Whereas for $h=6$ one may find a suitable variable
transformation, we did not succeed in this case. Nonetheless, the
solvability of the model may now be guaranteed now by relating the model to
gl$_{2}(\mathbb{R})\ltimes \mathbb{R}^{4}$, rather than gl(N), see (\ref{j1}%
) and (\ref{j2}).

Keeping from now on $\tilde{\kappa}_{8}=2\kappa _{8}$ and also $\kappa _{2}=1
$, we can bring the Hamiltonian into the desired form for the Schr\"{o}%
dinger operator (\ref{HJ}), (\ref{D})%
\begin{eqnarray}
\mathcal{D} &=&-J^{2}J^{1}-8J^{3}J^{1}-64\kappa _{8}J^{3}J^{8}+\left[
4(\gamma _{1}+\gamma _{2})-5-\frac{11}{9}\,\kappa _{8}\right] J^{1}  \notag
\\
&&-16\kappa _{8}\left[ (1-\gamma _{2})+\frac{4\kappa _{8}}{9}\right]
J^{8}+\gamma _{0}J^{2}+4\gamma _{0}J^{3}~.
\end{eqnarray}%
To turn this operator into the standard form (\ref{e2}) we follow the
procedure outlined in section 2. First, we compute 
\begin{equation}
G^{-1}=\det g^{-1}=-16I_{8}(I_{8}-4\kappa _{8}I_{2}^{4})~.  \label{det8}
\end{equation}%
Before we proceed further to compute the potential, let us see if we still
have a relation between $J$ and $G^{-1}$ of the type (\ref{JG}) for the
q-deformed algebra. In particular, we wish to see whether a relation of the
type (\ref{J2}) still holds. For this purpose we need first of all a notion
of positive q-deformed roots. We assume that these roots are generated in a
similar way as the ordinary roots, i.e. by repeated action of the Coxeter
element. Defining a q-deformed version of this\footnote{%
This q-deformed Coxeter element differs slightly from the one defined in 
\cite{q1,q2}, as here there is no $\tau $-transformation involved. The
Coxeter element in \cite{q1,q2} is only of order $h$ up to some factors of $q
$.} 
\begin{equation}
\sigma _{q}=\sigma _{2}^{q}\sigma _{1}^{q}  \label{Cox}
\end{equation}%
we compute the entire set of q-deformed roots $\Delta _{q}$ lying in the
orbits of the simple q-deformed roots $\alpha _{i}^{q}$, $\sigma
_{q}^{1}(\alpha _{i}^{q})$, $\sigma _{q}^{2}(\alpha _{i}^{q})$, $\ldots $ 
\begin{equation}
\begin{array}{||c|c|c||}
\hline
\sigma _{q}^{0} & \alpha _{1}^{q} & \alpha _{2}^{q} \\ \hline
\sigma _{q}^{1} & -\alpha _{3}^{q}=-(\alpha _{1}^{q}+\alpha _{2}^{q}) & 
\alpha _{6}^{q}=(2+\sqrt{2})\alpha _{1}^{q}+(1+\sqrt{2})\alpha _{2}^{q} \\ 
\hline
\sigma _{q}^{2} & ~-\alpha _{4}^{q}=-(1+\sqrt{2})\alpha _{1}^{q}-\sqrt{2}%
\alpha _{2}^{q}~~ & ~~\alpha _{7}^{q}=(2+2\sqrt{2})\alpha _{1}^{q}+(1+\sqrt{2%
})\alpha _{2}^{q}~~ \\ \hline
\sigma _{q}^{3} & -\alpha _{5}^{q}=-(1+\sqrt{2})\alpha _{1}^{q}-\alpha
_{2}^{q} & \alpha _{8}^{q}=(2+\sqrt{2})\alpha _{1}^{q}+\alpha _{2}^{q} \\ 
\hline
\sigma _{q}^{4} & -\alpha _{1}^{q} & -\alpha _{2}^{q} \\ \hline
\end{array}%
~.  \label{8roots}
\end{equation}%
Note that the order of the Coxeter element (\ref{Cox}) is indeed $h=8$, i.e. 
$\sigma _{q}^{8}=1$. We adopt now the same notion for positive and negative
roots as in the non-deformed case, that is we call $\alpha =\sum n_{i}\alpha
_{i}^{q}$ a positive root if all coefficients $n_{i}$ are positive. With
this notion the set of the $2h$ roots can be separated equally into $h$
positive and $h$ negative roots. We have verified this statement up to $h=20$
of even Coxeter numbers, which strongly suggests that it holds in general.
Using now (\ref{8roots}) we can compute the hyperplanes through the origin
to all positive q-deformed roots 
\begin{equation}
\begin{array}{ll}
p_{\alpha _{1}^{q}}=2x_{1}-(2+\sqrt{2})x_{2} & p_{\alpha
_{2}^{q}}=x_{1}-2x_{2} \\ 
p_{\alpha _{3}^{q}}=\sqrt{2}x_{1}-(2+\sqrt{2})x_{2}~~~~\qquad  & p_{\alpha
_{6}^{q}}=\sqrt{2}x_{1}-2(1+\sqrt{2})x_{2} \\ 
p_{\alpha _{4}^{q}}=x_{2} & p_{\alpha _{7}^{q}}=x_{1} \\ 
p_{\alpha _{5}^{q}}=x_{1}-x_{2} & p_{\alpha _{8}^{q}}=(2+\sqrt{2})x_{1}-2(1+%
\sqrt{2})x_{2}~~.%
\end{array}%
\end{equation}%
Comparing now with (\ref{det8}) we have once again a relation between $J$
and $G^{-1}$of the type (\ref{JG}) where $J$ can be expressed as a product
of hyperplanes (\ref{J2}) 
\begin{equation}
G^{-1}=\frac{\kappa _{8}^{2}}{4(3+2\sqrt{2})}\prod\limits_{\alpha ^{q}\in
\Delta _{q}^{+}}\left( p_{\alpha ^{q}}\right) ^{2}  \label{fact}
\end{equation}%
The two factors in (\ref{det8}) admit yet a further interpretation.
Organizing the roots into two sets $\Delta _{s}^{q}$ and $\Delta _{l}^{q}$
of short and long roots, respectively, we find the identities 
\begin{equation}
-\frac{\kappa _{8}}{4}\prod\limits_{\alpha ^{q}\in \Delta _{s}}\left(
p_{\alpha ^{q}}\right) ^{2}=I_{8}-4\kappa _{8}I_{2}^{4}\quad \text{and\quad }%
\frac{\kappa _{8}}{16(3+2\sqrt{2})}\prod\limits_{\alpha ^{q}\in \Delta
_{l}}\left( p_{\alpha ^{q}}\right) ^{2}=I_{8}~.
\end{equation}

\noindent According to (\ref{pre}) we make now the following ansatz for the
pre-potential 
\begin{equation}
P_{0}=e^{I_{2}},\qquad P_{1}=I_{8}-4\kappa _{8}I_{2}^{4},\qquad P_{2}=I_{8}.
\end{equation}%
\noindent From formula (\ref{pot}) and including also the $P_{0}$-term we
then compute the potential to

\begin{equation}
V=\frac{\gamma _{0}^{2}}{4}I_{2}-16\lambda _{1}\kappa _{8}\frac{I_{2}^{3}}{%
I_{8}-4\kappa _{8}I_{2}^{4}}+16\lambda _{2}\kappa _{8}\frac{I_{2}^{3}}{I_{8}}
\end{equation}%
with $\lambda _{i}=(\gamma _{i}^{2}-1/4)$ for $i=1,2$. Using the above
mentioned identities or directly (\ref{pott}), we can also re-write this
potential in terms of the $z$-variables. First of all we compute%
\begin{eqnarray}
P_{1} &=&\frac{1}{4^{2}3^{3}}\left( z_{1}-z_{3}\right) ^{2}\left(
z_{1}+z_{3}-2z_{3}\right) ^{2}\dprod\limits_{\varepsilon =\pm 1}\left[
(1+\varepsilon \sqrt{3})z_{1}+(1-\varepsilon \sqrt{3})z_{3}-2z_{2}\right]
^{2} \\
P_{2} &=&\frac{1}{4^{3}3^{4}}\dprod\limits_{\varepsilon ,\bar{\varepsilon}%
=\pm 1}\left[ (1-\bar{\varepsilon}\sqrt{3}-\varepsilon \sqrt{6})z_{1}+(1+%
\bar{\varepsilon}\sqrt{3}+\varepsilon \sqrt{6})z_{3}-2z_{2}\right] ^{2}~.
\end{eqnarray}%
We find 
\begin{eqnarray}
V &=&\frac{1}{2}\omega ^{2}\dsum\limits_{k=1}^{3}z_{k}^{2}+\frac{\lambda _{1}%
}{\left( z_{1}-z_{3}\right) ^{2}}+\frac{3\lambda _{1}}{\left(
z_{1}+z_{3}-2z_{2}\right) ^{2}}  \notag \\
&&+\dsum\limits_{\varepsilon =\pm 1}\frac{6\lambda _{1}}{\left[
(1+\varepsilon \sqrt{3})z_{1}+(1-\varepsilon \sqrt{3})z_{3}-2z_{2}\right]
^{2}} \\
&&+\dsum\limits_{\varepsilon ,\bar{\varepsilon}=\pm 1}\frac{6(2+\varepsilon 
\bar{\varepsilon}\sqrt{2})\lambda _{2}}{\left[ (1-\bar{\varepsilon}\sqrt{3}%
-\varepsilon \sqrt{6})z_{1}+(1+\bar{\varepsilon}\sqrt{3}+\varepsilon \sqrt{6}%
)z_{3}-2z_{2}\right] ^{2}}  \notag
\end{eqnarray}%
with $\omega =\gamma _{0}\sqrt{3}/(2\sqrt{2})$. We omitted here a constant
which contains the center of mass coordinate. Remarkably, all off-diagonal
terms, that is terms in (\ref{pot}) with $i\neq j$, cancel each other. This
potential has a very similar structure as the usual Calogero potentials (\ref%
{v2}), but it involves now deformed two and three-particle interactions. We
find similar structures for higher values of $h$. Responsible for this
structure is the fact that we can still factorize$\ J$, and therefore $%
G^{-1} $ in terms of products of hyperplanes as in (\ref{fact}). Remarkably,
these potentials are all exactly solvable by construction.

\subsection{Odd Coxeter numbers, h=5}

The structure for theories with odd values of the Coxeter number is somewhat
different. Let us consider $h=5$ in more detail. In that case the relation
for the deformation parameter (\ref{qq}) simply yields a root of unity 
\begin{equation}
q=e^{i\pi /10}\qquad \text{and\qquad }[3]_{q}=\frac{3+\sqrt{5}}{2}.
\end{equation}%
Replacing now in equation (\ref{Iq}) the variables $\vec{w}\rightarrow \vec{x%
}$, the invariant of degree $5$ in the $x$-variables is still not too
lengthy, unlike for greater values of $h$, and reads in this case 
\begin{equation}
I_{5}^{q}=\kappa _{5}\left( \frac{1}{2}(3-\sqrt{5}%
)x_{1}^{4}x_{2}-2x_{1}^{3}x_{2}^{2}+(1+\sqrt{5})x_{1}^{2}x_{2}^{3}-\frac{1}{2%
}(1+\sqrt{5})x_{1}x_{2}^{4}\right) ~.  \label{Iq5}
\end{equation}%
Obviously $I_{2}^{q}$ and $I_{5}^{q}$ are algebraically independent as one
is of even and the other of odd degree, respectively. As for the even case,
we can proceed and carry out in (\ref{Iq5}) the substitutions $\vec{x}%
\rightarrow \vec{y}\rightarrow \vec{z}$ such that the invariants are
expressed in terms of the Jacobian relative coordinates$~z_{i}$. In these
coordinates the inverse Riemannian (\ref{invg}) results to 
\begin{equation}
g_{ij}^{-1}=\sum_{k=1}^{3}\frac{\partial I_{i}}{\partial z_{k}}\frac{%
\partial I_{j}}{\partial z_{k}}=\kappa _{2}(5-\sqrt{5})\left( 
\begin{array}{cc}
\frac{1}{5}I_{2} & \frac{1}{2}I_{5} \\ 
\frac{1}{2}I_{5} & 5I_{2}^{4}\kappa _{5}^{2}/\kappa _{2}^{5}%
\end{array}%
\right) _{ij}~.  \label{25}
\end{equation}%
Once again we have the problem that $g_{ij}^{-1}$ is not of degree two in
the variables $I_{i}$. As for $h=8$ we can relate once more to the gl$_{2}(%
\mathbb{R})\ltimes \mathbb{R}^{5}$algebra, see (\ref{j1}) and (\ref{j2}).
From now on we keep $\kappa _{2}=1$ and bring the Hamiltonian into the
desired form (\ref{HJ}), (\ref{D})%
\begin{equation}
\mathcal{D}=(\sqrt{5}-5)\left( \frac{1}{5}\,J^{2}J^{1}+J^{1}J^{3}+5\kappa
_{5}^{2}J^{5}J^{9}-\left( \gamma _{1}-\frac{7}{10}-\frac{2\kappa }{3}\right)
J^{1}-\frac{\gamma _{0}}{5}J^{2}+\frac{\gamma _{0}}{2}\,J^{3}\right)
\end{equation}%
We proceed similarly as in the previous subsection and compute from (\ref{25}%
) 
\begin{equation}
G^{-1}=\det g^{-1}=\frac{5}{2}(\sqrt{5}-3)(I_{5}^{2}-4\kappa
_{5}^{2}I_{2}^{5})~.
\end{equation}%
In order to obtain Calogero type potentials it is vital to factorize $G^{-1}$
further into linear polynomials. Let us proceed analogously as for even
Coxeter numbers and compute the orbits of the q-deformed Coxeter element (%
\ref{Cox}). We find 
\begin{equation}
\begin{array}{||c|c|c||}
\hline
\sigma _{q}^{0} & \alpha _{1}^{q} & \alpha _{2}^{q} \\ \hline
\sigma _{q}^{1} & \alpha _{3}^{q}=-(\alpha _{1}^{q}+\alpha _{2}^{q}) & 
\alpha _{7}^{q}=\frac{1}{2}(3+\sqrt{5})\alpha _{1}^{q}+\frac{1}{2}(1+\sqrt{5}%
)\alpha _{2}^{q} \\ \hline
\sigma _{q}^{2} & ~\alpha _{4}^{q}=-\frac{1}{2}(1+\sqrt{5})\alpha _{1}^{q}-%
\frac{1}{2}(\sqrt{5}-1)\alpha _{2}^{q}~~ & \alpha _{8}^{q}=~~\frac{1}{2}(1+%
\sqrt{5})\alpha _{1}^{q}~~ \\ \hline
\sigma _{q}^{3} & \alpha _{5}^{q}=\frac{1}{2}(\sqrt{5}-1)\alpha _{2}^{q} & 
\alpha _{9}^{q}=-\frac{1}{2}(1+\sqrt{5})(\alpha _{1}^{q}+\alpha _{2}^{q}) \\ 
\hline
\sigma _{q}^{4} & \alpha _{6}^{q}=\frac{1}{2}(1+\sqrt{5})\alpha
_{1}^{q}+\alpha _{2}^{q} & \alpha _{10}^{q}=-\frac{1}{2}(3+\sqrt{5})\alpha
_{1}^{q}-\alpha _{2}^{q} \\ \hline
\sigma _{q}^{5} & \alpha _{1}^{q} & \alpha _{2}^{q} \\ \hline
\end{array}
\label{ro5}
\end{equation}%
Note that the order of the Coxeter element (\ref{Cox}) is still $h$, that is
in this case $\sigma _{q}^{5}=1$. For odd values of the Coxeter number we
can still separate the roots into positive and negative roots, but now the
negative roots can no longer be obtained by reversing the signs of all
positive roots. There are now positive roots without a negative counterpart.
Unfortunately, as a consequence of this the factorization property (\ref{J2}%
) does no longer hold in its stated form. Nonetheless, one can still select
some hyperplanes obtained from the root system (\ref{ro5}) and factorize $%
G^{-1}$, albeit now the selection principle does no longer favour the
positive roots and is less clear. We compute 
\begin{equation}
\begin{array}{ll}
p_{\alpha _{1}^{q}}=x_{1}-\frac{1}{4}(3+\sqrt{5})x_{2} & p_{\alpha
_{2}^{q}}=x_{1}-2x_{2} \\ 
p_{\alpha _{3}^{q}}=x_{1}-(2+\sqrt{5})x_{2}~~~~ & p_{\alpha _{7}^{q}}=x_{1}+%
\frac{1}{2}(1+\sqrt{5})x_{2} \\ 
p_{\alpha _{4}^{q}}=\frac{1}{2}(1+\sqrt{5})x_{1}-x_{2}\qquad ~~~~ & 
p_{\alpha _{8}^{q}}=x_{1}-\frac{1}{4}(3+\sqrt{5})x_{2} \\ 
p_{\alpha _{5}^{q}}=x_{1}-2x_{2} & p_{\alpha _{9}^{q}}=x_{1}+\frac{1}{2}(1-%
\sqrt{5})x_{2} \\ 
p_{\alpha _{6}^{q}}=\frac{1}{2}(\sqrt{5}-1)x_{1}+x_{2} & p_{\alpha
_{10}^{q}}=x_{1}-\frac{1}{2}(\sqrt{5}-1)x_{2}%
\end{array}%
\end{equation}%
and construct from this 
\begin{equation}
J=p_{\alpha _{1}^{q}}p_{\alpha _{2}^{q}}p_{\alpha _{3}^{q}}p_{\alpha
_{7}^{q}}p_{\alpha _{10}^{q}}~.
\end{equation}%
Now, unlike as in the even case, the splitting into long and short roots no
longer corresponds to factors in terms of hyperplanes.

\noindent For the pre-potential we make now the ansatz 
\begin{equation}
P_{0}=e^{I_{2}},\qquad P_{1}=I_{5}^{2}-4\kappa _{5}^{2}I_{2}^{5}.
\end{equation}%
With formula (\ref{pot}) we then compute the potential in terms of invariant
polynomials to

\begin{equation}
V=\frac{\gamma _{0}^{2}}{4}\left( 1-\frac{1}{\sqrt{5}}\right) I_{2}+5(\sqrt{5%
}-5)\lambda \kappa _{5}^{2}\frac{I_{2}^{4}}{I_{5}^{2}-4\kappa
_{5}^{2}I_{2}^{5}},  \label{vh5}
\end{equation}%
with $\lambda =(\gamma _{1}^{2}-1/4)$. Once again we can also re-write $V$
in terms of the $z$-variables. First we factorize 
\begin{equation}
P_{1}=\frac{5^{2}}{3^{4}4^{5}}(z_{1}-z_{3})^{2}\dprod\limits_{\bar{%
\varepsilon},\varepsilon =\pm 1}\left[ (1+\sqrt{3+\frac{6\varepsilon }{\sqrt{%
5}}})z_{2+\bar{\varepsilon}}+(1-\sqrt{3+\frac{6\varepsilon }{\sqrt{5}}})z_{2-%
\bar{\varepsilon}}-2z_{2}\right] ^{2}
\end{equation}%
from which we deduce the potential to%
\begin{equation}
V=\frac{1}{2}\omega ^{2}\dsum\limits_{k=1}^{3}z_{k}^{2}+\frac{\lambda }{%
(z_{1}-z_{3})^{2}}+\dsum\limits_{\bar{\varepsilon},\varepsilon =\pm 1}\frac{%
6(1+\varepsilon /\sqrt{5})\lambda }{\left[ (1+\sqrt{3+\frac{6\varepsilon }{%
\sqrt{5}}})z_{2+\bar{\varepsilon}}+(1-\sqrt{3+\frac{6\varepsilon }{\sqrt{5}}}%
)z_{2-\bar{\varepsilon}}-2z_{2}\right] ^{2}}
\end{equation}%
with $\omega ^{2}=3(3-\sqrt{5})\gamma _{0}^{2}/20$. Once again all
off-diagonal terms cancel each other and as in the even case this potential
is also of Calogero type. We find similar types of potentials for higher
values of the Coxeter number.

\section{Conclusions}

It has been shown previously that solvability of certain types of
Hamiltonians can be established \cite{Tur0,RT,Cap,TurF4} by relating the
differential operators inside the Hamiltonians, that is essentially the
Laplace operator, to a representation of the gl(N)-Lie algebra. This
formulation can be made very systematic by associating the differential
structure to polynomial invariants of Coxeter groups. This led the authors
of \cite{HR1,HR2,HR3,HR4} to propose a procedure which allows to construct
solvable Hamiltonians by taking the structure of the polynomial invariants
as a starting point. Here we showed that this procedure can be extended
successfully to polynomial invariants of q-deformed Coxeter groups. We
constructed some potentials resulting from these type of invariants. Due to
the fact that the Jacobian determinant can still be factorized in terms of
linear polynomials the resulting potentials are of Calogero type.

There are several open issues and unanswered questions which deserve further
investigations. Clearly it would be interesting to carry out the outlined
procedure explicitly for other algebras than $G_{2}^{q}$. The presented
example indicates that one can expect similar structures beyond $G_{2}^{q}$.
Eventually one should aim at a unified formulation, as opposed to
case-by-case studies, analogously to the non-deformed case as indicated in
section 5. Crucial will be here the factorization of the Jacobian
determinant $J$. In the presented example this works nicely for even Coxeter
numbers, but for odd $h$ the example hints that one possibly has to employ a
different q-deformed Coxeter transformation in order to obtain a definite
criterion for the selection of the hyperplanes which yields the
factorization of $J$.

To achieve a unified formulation it will be important to have systematic and
generic expressions for the polynomial invariants \cite{FK}. In the above
analysis we have seen that the choice of a suitable basis is absolutely
crucial for this task. The favoured one is the eigenbasis of the Coxeter
element as we have demonstrated.

Since we have shown that one can extend the approach from Coxeter to
q-deformed Coxeter groups, it is also natural to suspect that one might as
well employ it for reflection groups the general type introduced in \cite%
{Zuber1}.

Naturally it appears also possible to construct potential of Sutherland type
by using different types of coordinates \cite{FK}.

To find the explicit wavefunction for the above Hamiltonians is now also an
obvious question to ask. Following the quoted literature there is a
straightforward procedure to construct them from the above mentioned
results. In this context one might also address technical questions as for
instance self-adjointness similar as it has been done in Calogero's original
work for hardcore boundary conditons \cite{Cal1,Cal2,Cal3}. For slightly
more general boundary conditions, see e.g. \cite{Calc}. Conceptionally one
should stress that solvability in the sense provided here is far more
constructive with regard to this question than integrability. The latter
usually just guarantees the existence of exact solutions, whereas
solvability is already tied closely to the explicit solutions.

\ 

\noindent \textbf{Acknowledgments: } We are grateful to R. Twarock for
bringing this subject to our attention and H.W. Braden, A. Cox for useful
comments. C.K. is supported by EPSRC Grant/R93773/01 and thanks City
University for kind hospitality.


\begin{thebibliography}{99}
\bibitem{Cal1} F.~Calogero, \newblock Ground state of one-dimensional N body
system, \newblock J. Math. Phys. \textbf{10}, 2197--2200 (1969).

\bibitem{Cal2} F.~Calogero, \newblock Solution of a three-body problem in
one-dimension, \newblock J. Math. Phys. \textbf{10}, 2191--2196 (1969).

\bibitem{Cal3} F.~Calogero, \newblock Solution of the one-dimensional N body
problems with quadratic and/or  inversely quadratic pair potentials, %
\newblock J. Math. Phys. \textbf{12}, 419--436 (1971).

\bibitem{Suth1} B.~Sutherland, \newblock Quantum many body problem in
one-dimension: Ground state, \newblock J. Math. Phys. \textbf{12}, 246--250
(1971).

\bibitem{Suth2} B.~Sutherland, \newblock Quantum many body problem in
one-dimension, \newblock J. Math. Phys. \textbf{12}, 251--256 (1971).

\bibitem{Suth3} B.~Sutherland, \newblock Exact results for a quantum many
body problem in one- dimension, \newblock Phys. Rev. \textbf{A4}, 2019--2021
(1971).

\bibitem{Suth4} B.~Sutherland, \newblock Exact results for a quantum many
body problem in one- dimension. 2, \newblock Phys. Rev. \textbf{A5},
1372--1376 (1972).

\bibitem{OP1} M.~A. Olshanetsky and A.~M. Perelomov, \newblock Quantum
completely integrable systems connected with semisimple Lie  algebras, %
\newblock Lett. Math. Phys. \textbf{2}, 7--13 (1977).

\bibitem{OP2} M.~A. Olshanetsky and A.~M. Perelomov, \newblock Classical
integrable finite dimensional systems related to Lie algebras, \newblock %
Phys. Rept. \textbf{71}, 313-400 (1981).

\bibitem{OP3} M.~A. Olshanetsky and A.~M. Perelomov, \newblock Quantum
integrable systems related to Lie algebras, \newblock Phys. Rept. \textbf{94}%
, 313--404 (1983).

\bibitem{DIO} D.~I. Olive and N.~Turok, \newblock Local conserved densities
and zero curvature conditions for Toda lattice field theories, \newblock %
Nucl. Phys. \textbf{B257}, 277-301 (1985).

\bibitem{Sas6} A.~J. Bordner, E.~Corrigan, and R.~Sasaki, \newblock %
Calogero-Moser models. I: A new formulation, \newblock Prog. Theor. Phys. 
\textbf{100}, 1107--1129 (1998).

\bibitem{Sas5} A.~J. Bordner, E.~Corrigan, and R.~Sasaki, \newblock %
Generalised Calogero-Moser models and universal Lax pair operators, %
\newblock Prog. Theor. Phys. \textbf{102}, 499--529 (1999).

\bibitem{Sas4} S.~P. Khastgir, A.~J. Pocklington, and R.~Sasaki, \newblock %
Quantum Calogero-Moser Models: Integrability for all Root Systems, \newblock %
J. Phys. \textbf{A33}, 9033--9064 (2000).

\bibitem{Sas3} V.~I. Inozemtsev and R.~Sasaki, \newblock Universal Lax pairs
for Spin Calogero-Moser Models and Spin Exchange  Models, \newblock J. Phys. 
\textbf{A34}, 7621--7632 (2001).

\bibitem{Sas2} E.~Corrigan and R.~Sasaki, \newblock Quantum vs classical
integrability in Calogero-Moser systems, \newblock J. Phys. \textbf{A35},
7017--7062 (2002).

\bibitem{Tur0} A.~Turbiner, \newblock Lie algebras and linear operators with
invariant subspaces, \newblock Lie Algebras, Cohomologies and New Findings
in Quantum Mechanics,  Contemp. Math. AMS, (eds N. Kamran and P.J. Olver) 
\textbf{160}, 263--310  (1994).

\bibitem{RT} W.~R{\"u}hl and A.~Turbiner, \newblock Exact solvability of the
Calogero and Sutherland models, \newblock Mod. Phys. Lett. \textbf{A10},
2213--2222 (1995).

\bibitem{Cap} A.~Capella, M.~Rosenbaum, and A.~Turbiner, \newblock %
Solvability of the $G_2$ integrable system, \newblock Int. J. Mod. Phys. 
\textbf{A13}, 3885--3904 (1998).

\bibitem{TurF4} K.~G. Boreskov, J.~C. Lopez~Vieyra, and A.~V. Turbiner, %
\newblock Solvability of the $F_{4}$ integrable system, \newblock Int. J.
Mod. Phys. \textbf{A16}, 4769-4801 (2001).

\bibitem{Brink} L.~Brink, A.~Turbiner, and N.~Wyllard, \newblock Hidden
Algebras of the (super) Calogero and Sutherland models, \newblock J. Math.
Phys. \textbf{39}, 1285--1315 (1998).

\bibitem{HR1} O.~Haschke and W.~R{\"u}hl, \newblock Exactly solvable
dynamical systems in the neighborhood of the  Calogero model, \newblock Int.
J. Mod. Phys. \textbf{A14}, 387--408 (1999).

\bibitem{HR2} O.~Haschke and W.~R{\"u}hl, \newblock Exactly solvable quantum
models of Calogero and Sutherland type with  translation invariant
four-particle interactions, hep-th/9807194, \newblock (1998).

\bibitem{HR3} O.~Haschke and W.~R{\"u}hl, \newblock Is it possible to
construct exactly solvable models?, \newblock Lect. Notes Phys. \textbf{539}%
, 118--140 (2000).

\bibitem{HR4} O.~Haschke and W.~R{\"u}hl, \newblock An exactly solvable
model of the Calogero type for the icosahedral  group, \newblock Mod. Phys.
Lett. \textbf{A13}, 3109--3122 (1998).

\bibitem{Reidun} R.~Twarock, \newblock An exactly solvable Calogero model
for a non-integrally laced group, \newblock Phys. Lett. \textbf{A275},
169--172 (2000).

\bibitem{q2} A.~Fring, C.~Korff, and B.~J. Schulz, \newblock On the
universal representation of the scattering matrix of affine  Toda field
theory, \newblock Nucl. Phys. \textbf{B567}, 409--453 (2000).

\bibitem{FK} A.~Fring and C.~Korff, \newblock In preparation.

\bibitem{Hum2} J.~E. Humphreys, \newblock Reflection Groups and Coxeter
Groups, \newblock Cambridge University Press, Cambridge (1990).

\bibitem{q1} T.~Oota, \newblock q-deformed Coxeter element in non-simply
laced affine Toda field  theories, \newblock Nucl. Phys. \textbf{B504},
738--752 (1997).

\bibitem{Hum} J.~E. Humphreys, \newblock Introduction to Lie Algebras and
Representation Theory, \newblock Springer, Berlin (1972).

\bibitem{Symm} I.~G. Macdonald, \newblock Symmetric Functions and Hall
Polynomials, \newblock Clarendon Press, Oxford (1979).

\bibitem{Wolf} J.~Wolfes, \newblock On the three-body linear problem with
three body interaction, \newblock J. Math. Phys. \textbf{15}, 1420--1424
(1974).

\bibitem{Zuber1} J.~B. Zuber, \newblock Graphs and reflection groups, %
\newblock Commun. Math. Phys. \textbf{179}, 265--294 (1996).

\bibitem{Calc} B.~Basu-Mallick, P.~K. Ghosh, and K.~S. Gupta, \newblock %
Inequivalent Quantizations of the Rational Calogero Model, \newblock Phys.
Lett. \textbf{A311}, 87--92 (2003).
\end{thebibliography}
\end{document}